\documentclass[twocolumn,superscriptaddress,amssymb,amsmath,nobibnotes,aps,prd]{revtex4}

\def\bec{\begin{center}}
\def\eec{\end{center}}
\def\beq{\begin{equation}}
\def\eeq{\end{equation}}
\def\bseq{\begin{subequations}}
\def\eseq{\end{subequations}}
\def\bea{\begin{eqnarray}}
\def\eea{\end{eqnarray}}

\usepackage{xcolor}
\usepackage{comment}
\usepackage{graphicx}
\usepackage{tensor}
\usepackage{appendix}

\usepackage[linktocpage]{hyperref}
\hypersetup{colorlinks=true,
citecolor=magenta, linkcolor=blue}

\usepackage[caption=false]{subfig}

\begin{document}

\title{ Relativistic dynamics of moving mirrors in CFT$_2$: \\ quantum backreaction and black holes. }

\author{Piyush Kumar}
\email{kumar@uni-wuppertal.de}
\affiliation{Department of Physics, Wuppertal University, Gaussstr. 20, D-42119, Wuppertal, Germany}
\author{Ignacio A. Reyes}
\email{ireyesraffo@gmail.com}
\affiliation{Institute for Theoretical Physics, University of Amsterdam, Amsterdam, 1098 XH, The Netherlands}
\author{Jakob Wintergerst}
\email{jakobw@physik.hu-berlin.de}
\affiliation{Institute of Physics, Humboldt University Berlin, Zum großen Windkanal 6, 12489 Berlin, Germany}

\date{\today}

\begin{abstract}

There is a well-known correspondence between the physics of black hole evaporation and that of moving mirrors in QFT. However, most analyses in this subject rely on prescribed mirror trajectories. Here, we study the flat-space dynamics of $1+1$-dimensional Conformal Field Theories interacting with a relativistic boundary particle of mass $m$ acting as a perfect mirror. The trajectory of the latter is not fixed but follows its own relativistic equation of motion $F^\mu=ma^\mu$. For given initial conditions at past null infinity, we find the boundary particle's trajectory and the reflected energy-momentum of the quantum fields. For incoming vacuum states, the solution yields mirror orbits that correspond to extremal black holes. For the class of incoming states that produce orbits becoming null in finite proper time -- corresponding to the formation of a horizon -- at the classical level, the quantum backreaction avoids this endpoint rendering the mirror's velocity in lightcone coordinates finite. We investigate the behavior of the Averaged Null Energy Condition, which in this setup reduces to a boundary term.

\end{abstract}

\maketitle


\section{Introduction.}

One of the most surprising features that distinguish quantum theories from their classical counterparts is the existence of zero-point energies. Although not directly measurable, their differences can create real forces such as the celebrated static Casimir effect\,\cite{Casimir:1948dh}. A major generalization of this phenomenon was found by Moore\,\cite{1970JMP....11.2679M}, who considered a quantized field contained within a perfectly reflecting cavity with \textit{moving} boundaries, showing that in addition to creating a force, an accelerating mirror will produce radiation. This is known as the Dynamical Casimir effect. 

The connection between moving mirrors and black holes was pioneered by Fulling and Davies\,\cite{Davies:1977yv}, who showed how to map this problem to that of a collapsing star in GR as in Hawking's setup \cite{Hawking:1975vcx}. They showed that for any mirror trajectory that becomes asymptotically null as
\begin{align}\label{FD}
    x^+(x^-)=x_0^+-\beta e^{-x^-/\beta},
\end{align}
where $x^\pm=t\pm x$, the outgoing quantum state matches precisely with that of the Hawking effect at late times for a black hole at temperature $(2\pi \beta)^{-1}$. 

\begin{figure}[h]
\subfloat[\label{A}]{\includegraphics[width=0.28\textwidth]{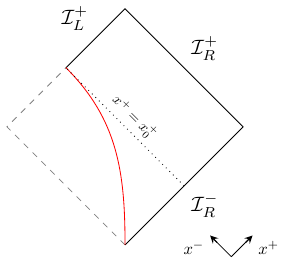}} \hfill
\subfloat[\label{B}]{\includegraphics[width=0.17\textwidth]{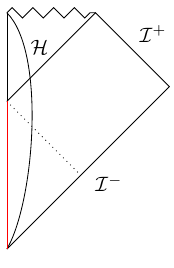}}
\caption{(a): Penrose diagram of a moving mirror (in red), that smoothly departs from its inertial trajectory in the past to become null asymptotically according to \eqref{FD}, approaching the finite value $x_0^+$. (b): Penrose diagram of a four-dimensional spherically symmetric black hole formed by collapsing matter. The curve represents the surface of the collapsing star. After it crosses its Schwarzschild radius, an event horizon is formed, which in this diagram is represented by $\mathcal{H}$. An outside observer with a constant Schwarzschild radius will see the origin (in red) recede away at late times precisely according to the trajectory on the left.}
\label{mirror:collapsingBH}
\end{figure}

In Fig. \ref{mirror:collapsingBH} we show a $1+1$ dimensional graphic representation of this correspondence, with the conformal diagram of the mirror moving along the trajectory \eqref{FD} on the left, and the corresponding one for gravitational collapse on the right. The two systems share many features. Prepared in the initial vacuum at $\mathcal{I}^-_R (\mathcal{I}^-)$, an observer at $\mathcal{I}_R^+ (\mathcal{I}^+)$ will perceive a thermal state for late times. The origin of the radiation in the mirror system is the accelerated boundary, whereas in the collapsing star, it is the rapidly changing gravitational field. 

The large advantage of moving mirrors is that one can study radiative effects of black holes while avoiding the complications associated with the curvature of spacetime and the corresponding non-linearities.

Most of the discussion on moving mirrors in QFT -- including the `dynamical' Casimir effect  -- involves only mirror trajectories that are `prescribed', i.e. worldlines that are specified a priori and are not the outcome of some underlying dynamics. In this paper, we wish to make progress in addressing the truly dynamical version of this. We study the backreaction problem of a quantum field interacting with a classical boundary particle obeying its own equation of motion. The distinguishing features of our work is that our approach for the backreaction is fully relativistic, and valid for any CFT$_2$. 

We will only be concerned with perfectly reflecting classical (non-quantum mechanical) mirrors. Although real mirrors must become partially transparent at high enough frequencies, this does not seem relevant for the connection to gravitational collapse, as the mirror trajectory is mapped to the origin of coordinates of the contracting sphere. 

As is well known from classical electrodynamics, the radiation-reaction problem for perfectly reflecting mirrors suffers from certain pathologies. This is an old problem and we make no attempt to solve it. Rather, our main concern will be to understand the response of the mirror to the class of incoming energy-momenta associated with gravitational collapse in the gravity picture, with a particular interest in the behavior of the system just before a horizon would form.

The paper is organized as follows. In section \ref{sec:setup} we introduce the system and derive the equations of motion, first for classical and then for quantum fields. In section \ref{sec:vacuum} we study the vacuum solutions, i.e. those with zero incoming momentum from past null infinity. Section \ref{sec:singular} contains the main results of this paper. We consider the class of incoming stress tensors that are associated with horizon formation in the gravity picture mentioned above, solve for the quantum backreacted equations of motion, and study the resulting trajectories. We identify the class of incoming data for which the introduction of the conformal anomaly changes the causal structure qualitatively, preventing the formation of a `horizon'. In section \ref{sec:ANEC} we reexamine our results in light of the Averaged Null Energy Condition (ANEC) and its violations. We end with a summary and outlook in section \ref{sec:summary}. 

We use units where $c=\hbar=1$.

\section{Fields Coupled to a Boundary Particle.}\label{sec:setup}

An elegant approach to the interaction between a classical relativistic particle acting as a mirror for a QFT in $1+1$ dimensions was employed by Chung-Verlinde \cite{Chung:1993rf}.

Let us begin by considering the problem of a boundary particle with mass $m$ interacting with a classical field. We impose perfectly reflecting boundary conditions and restrict the field to exist only to the right of the boundary. Upon reflection, momentum is transferred from the field to the particle. The worldline of the particle is parameterized in terms of proper time $x^{\mu}(\tau)$. The dynamics are governed by Newton's equation
\begin{align}
    F^\mu=m \frac{d^2 x^{\mu}}{d\tau^2},
\end{align}
which, in terms of the null coordinates $x^{\pm}=t\pm x$, reads
\begin{align}
    F_{\pm}=\frac{m}{2} \frac{d^2 x^{\mp}}{d\tau^2}. 
\end{align}
In the classical theory, energy flux is reflected according to
\begin{align}
    T_{--}(dx^-)^2=T_{++}(dx^+)^2.
    \label{reflection:classical}
\end{align}
To determine the force components $F_{\pm}$, we consider the proper time normalization condition
\begin{align}
\dot{x}^+\dot{x}^-=1,
\label{norm:cond}
\end{align}
where an overdot represents differentiation with respect to proper time. It follows that
\begin{align}\label{force:perp}
F_+ \dot{x}^+ + F_- \dot{x}^-=0.
\end{align}
By comparing Eq. \eqref{reflection:classical} and \eqref{force:perp}, we can infer the classical equations of motion for the boundary particle
\begin{align}\label{eom:classical}
\frac{m}{2}\ddot{x}^{\pm}=\mp T_{\mp \mp} \dot{x}^{\mp}.
\end{align}
\subsection{Anomalous Equations of Motion.}
Upon quantization, the classical fields are promoted to operators. In computations, the corresponding expressions are replaced by their expectation values. As the stress tensor is quadratic in the field, its expectation value is divergent and has to be regularized, for example by means of point-splitting.  This explicit use of coordinates breaks conformal symmetry and the reflected stress tensor attains an anomalous term
\begin{align}
    T_{--}=&\left(\frac{dx^+}{dx^-}\right)^{2} \left[T_{++}+\frac{c}{24 \pi} \{x^-,x^+\} \right]
    \label{reflection:quantum},
\end{align}
where the brackets $\{ \ ,\ \}$ denote the Schwarzian derivative and $c$ is the central charge of the CFT. We will now derive the quantum analog of (\ref{eom:classical}), by rewriting the quantum reflection equation (\ref{reflection:quantum}) in a similar form as \eqref{force:perp} to read off the respective force components. This method is not unique; here we will proceed by casting the transformation law in a symmetric form, which leads to cancellations that simplify the equations of motion. We can write (\ref{reflection:quantum}) as
\begin{align}
\begin{split}
    &\left[T_{--}+\frac{c}{48 \pi}\{x^+,x^-\}\right](\dot{x}^-)^2 \\
    &=\left[T_{++}+\frac{c}{48 \pi}\{x^-,x^+\} \right] (\dot{x}^+)^2.
\end{split}
\label{refl:symm}
\end{align}
Expressing the Schwarzian derivative in terms  of proper time gives
\begin{align}\label{schwartzian}
    \{x^-,x^+\}=2 \left(\dddot{x}^- \dot{x}^--(\ddot{x}^-)^2\right),
\end{align}
with a similar expression for $\{x^+,x^-\}$.
When inserted into (\ref{refl:symm}), the $(\ddot{x}^{\pm})^2$ terms cancel and the quantum equations of motion follow
\begin{align}\label{aeom1}
     \frac{m}{2}\dot{v}^{\pm}\pm   \frac{T_{\mp \mp}}{v^{\pm}} \pm \frac{c}{24 \pi} \ddot{v}^{\pm}=0,
\end{align}
where $v^{\pm}=\dot{x}^{\pm}$. These differential equations are of third order in time. This is typical for systems that account for back-reaction effects of acceleration-induced radiation. A canonical example that exemplifies this behavior is the radiation reaction experienced by an accelerated charged point particle within the framework of classical electrodynamics.

Since \eqref{aeom1} depends only on the velocity and its derivatives in a very simple way, we can integrate once to obtain
\begin{align}\label{aeom2}
     {v}^\pm \pm P_{\mp}\pm q \dot{v}^\pm=A^{\pm},
\end{align}
with $A^{\pm}$  constants of integration. In \eqref{aeom2} we have defined 
\begin{align}
    q=\frac{c}{12m\pi},
\end{align}
which has dimensions of length, and the integrated momentum
\begin{align}\label{total:momentum}
    P_{\pm}(x^\pm)=\frac{2}{m}\int_{-\infty}^{x^{\pm}(\tau)} d\tilde{x}^{\pm} T_{\pm \pm}(\tilde x^\pm).
\end{align}
The case $q=0$ corresponds to the previously discussed classical system. The two components of \eqref{aeom1} or \eqref{aeom2} are not independent but are related to each other via the proper time normalization condition (\ref{norm:cond}). When specifying the quantum state at $\mathcal{I}^-$, as we shall do in the following, $P_+$ represents initial data, while $P_-$ is determined by the mirror trajectory. Thus, in this case, it is simpler to solve the equation corresponding to $x^-$.

We can express the equations of motion also in terms of the coordinates, eliminating proper time. Parameterizing the mirror trajectory as $x^-=f(x^+)$, Eq. \eqref{aeom2} can be written as 
\begin{align}\label{aeom3}
    \sqrt{f'}-\frac{q}{2} \frac{f''}{f'}-P_+=A^-.
\end{align}

To summarize, there are two equivalent ways of describing the motion of the particle: Eq. \eqref{aeom2}, which uses proper time to parameterize the two coordinate functions $x^\pm(\tau)$, and \eqref{aeom3} where we eliminate proper time in favor of a single function $x^-=f(x^+)$. Either description proves more useful in different contexts: In the following section, we will analyze the solutions when the incoming state is the vacuum, hence $P_+=0$. There it is natural to use the simpler parameterization in terms of proper time \eqref{aeom2}. However, as soon as we introduce an external momentum in section \ref{sec:singular}, proper time is no longer a convenient parameter so we will use \eqref{aeom3}.
\section{Vacuum Solutions.}\label{sec:vacuum}
\subsection{Classical}
We begin by solving the classical equations of motion for the incoming vacuum, i.e. $P_{+}=0$. Obtained by setting $q=0$ in \eqref{aeom2}, the solutions are trivial
\begin{align}
    v^{\pm}=A^{\pm},
\end{align}
with inertial motion being the only solution. As expected, there is no radiation.
\subsection{Quantum}
The corresponding quantum equation reads
\begin{align}
    v^--q \dot{v}^-=A^-\equiv A.
\end{align}
This equation can be readily integrated, resulting in
\begin{align}\label{vac:vel}
    v^-(\tau)=A+ B e^{\tau/q},
\end{align}
where $B$ is another integration constant. By integrating $v^-$ and the inverse expression $v^+=1/v^-$, we obtain the coordinates as functions of proper time
\begin{align}
    x^+(\tau)&=-\frac{q}{A}\log \left(A e^{-\tau/q}+B\right),\\
    x^-(\tau)&=A \tau+B q e^{\tau/q} \label{xm(tau)},
\end{align}
where we have set the integration constants to zero for simplicity. 

\begin{figure}[h]
\renewcommand\thesubfigure{\roman{subfigure}}
\subfloat[$A>0, B>0$\label{A}]{\includegraphics[width=0.24\textwidth]{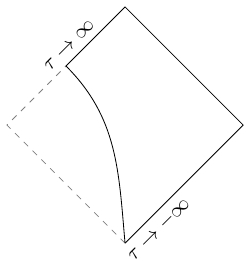}} \hfill
\subfloat[$A=0, B>0$ \label{B}]{\includegraphics[width=0.24\textwidth]{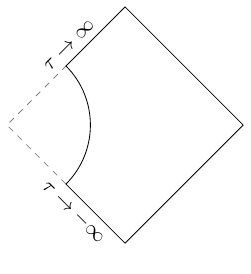}}\\
\subfloat[$A<0, B>0$\label{C}]{\includegraphics[width=0.24\textwidth]{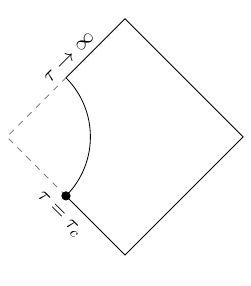}} \hfill
\subfloat[$A>0, B<0$\label{D}]{\includegraphics[width=0.24\textwidth]{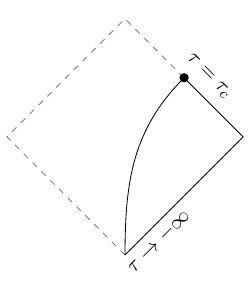}}
\caption{Conformal diagram of the four classes of vacuum solutions. The quantum field exists only to the right of the mirror. All future null left orbits have constant proper acceleration.  For orbits (iii) and (iv) there is a finite proper time along the worldline (indicated by the black dot) in the past and future, respectively. }
\label{VacuumPenrose}
\end{figure}

For $B=0$ the solutions again are inertial motion, which can be easily seen from \eqref{vac:vel}. In contrast to the classical case, however, we also have non-trivial solutions, corresponding to $B\neq0$. To examine the space of physical solutions, we note that future-oriented curves satisfy $v^\pm >0$, resulting in four classes of vacuum trajectories depending on the values of the integration constants $A$ and $B$, as shown in Fig \ref{VacuumPenrose}.

Even though the incoming state is the vacuum, the outgoing state in general is not. The reflected stress tensor is determined by \eqref{reflection:quantum}, which in terms of proper time can be compactly written as $T_{--}=-c \dot{\alpha}/(12 \pi (v^-)^2)$, where
\begin{align}
    \alpha(\tau)=-\frac{\dot{v}^-}{v^-}
\end{align}
is the proper acceleration. Substituting \eqref{xm(tau)} into this expression yields the reflected energy as a function of proper time along the worldline, 
\begin{align}
    T_{--}(\tau)=\frac{c A B e^{\tau /q}}{12 \pi q^2 \left(A+B e^{\tau /q}\right)^4}.
\end{align}

Orbits (i)-(iii) share an interesting property: in their asymptotic future the proper acceleration becomes constant,
\begin{align}
    \lim_{\tau \to \infty} \alpha(\tau)=-\frac{1}{q}\,.
\end{align}

Thus, although there is no radiation at late times, the causal structure in the future still resembles that of the black hole. Indeed it has been recently shown that trajectories of constant proper acceleration $-1/q$ correspond to extremal Reissner-N\"ordstrom black holes of mass $M=\frac{q}{2G}$ \,\cite{Good_2020}. This means that the extremal black hole that corresponds to our vacuum solutions has mass
\begin{align}
    M=\frac{c}{24\pi} \frac{1}{G m},
\end{align}
which is directly proportional to the number of degrees of freedom in the CFT, and inversely proportional to the mass of the boundary particle.

A crucial difference among the trajectories lies in the range of proper time $\tau$. Orbits (i) and (ii) have an infinite range of $\tau$ to both past and future. On the other hand, both (iii) and (iv) possess a critical time $\tau_c$, i.e., measured from any point along the trajectory, the particle's velocity becomes singular in finite proper time towards the past or future, respectively. This critical time is given by
\begin{align}
    \tau_c = q\log \left(-A/B\right)\,.
\end{align}
Interestingly, the radiated energy is strictly negative for the trajectories (iii) and (iv), where $A$ and $B$ have a relative sign. In section \ref{sec:ANEC} we will study the averaged null energy condition and return to those solutions and their negative energy flux.

As explained above, the alternative representation \eqref{aeom3} eliminates proper time. For completeness, we also provide the vacuum solutions for $x^-$ as a function of the coordinate $x^+$, 
\begin{align}
    \label{vac:coordinates}
    \begin{split}
    x^-=f(x^+)=&A^2 x^++\frac{q A}{1- Be^{A x^+/q}} \\
    &-q A\log \left(A- A B e^{A x^+/q}\right). 
     \end{split}
\end{align}
In the sense explained above, trajectories (i)-(iii) have a `horizon' located at $x^+_H= q/A \log \left(1/B\right)$.
In order to understand the asymptotic behavior of the mirror, we expand $x^+ \approx x^+_H$ which yields for those orbits
\begin{align}\label{moeb}
    f(x^+)\approx &\frac{q^2}{(x_H^+-x^+)}\,.
\end{align}
The map \eqref{moeb} is a M\"obius transformation, with vanishing Schwarzian derivative, which shows again the correspondence to extremal black holes.   

In terms of $x^-$ the outgoing energy flux is given by
\begin{align}
   T_{--}(x^-)= \frac{c  W\left(B/A e^{x^-/A q}\right)}{12 \pi A^2 q^2 \left(W\left(B/A e^{x^-/A q}\right)+1\right)^4},
\end{align}
where $W(x)$ is the Lambert function.

\section{Fulling-Davies Stress Tensors and Backreaction.}\label{sec:singular}
As explained above in the introduction, the problem of the gravitational collapse of a star towards its Schwarzschild radius is closely connected to mirror worldlines that in the asymptotic future follow the Fulling-Davies (FD) trajectory
\begin{align}\label{FD2}
    x^+(x^-)=x_0^+-\beta e^{-x^-/\beta},
\end{align}
where $(2\pi \beta)^{-1}$ is the associated Hawking temperature. The proper time  $d\tau=\sqrt{dx^+/dx^-}dx^-$ along  this trajectory yields
\begin{align}
    \tau=2\beta \left( 1- e^{-\frac{x^-}{2\beta}} \right) \underset{x^-\to \infty}{\to} 2\beta,
\end{align}
where we set $\tau=0$ at $x_0^-=0$. Thus the proper time towards the future is finite and determined by the associated black hole temperature.

Let us consider this analogy between gravitational collapse and accelerating mirrors more carefully. Take first the purely classical level. In GR, the `classical' (i.e. non-anomalous) source could be a fluid obeying a given equation of state. One then solves $G_{\mu\nu}=8\pi G T_{\mu\nu}$ which provides the classical background geometry. The counterpart in the moving mirror setup  corresponds to a classical field theory interacting with a boundary reflecting particle, obeying the classical equation of motion $ \frac{m}{2}\ddot{x}^{\pm}=\mp  T_{\mp \mp} \dot{x}^{\mp}$ as explained in \eqref{eom:classical}. Both the gravity and the mirror equations relate a `geometric' second-order operator to a source that depends on the energy-momentum of the fields. 

In the next level of approximation, we incorporate the effect of the QFT stress tensor as an additional source to the equations of motion of the geometry. In the gravitational context, this corresponds to solving the problem of gravitational collapse by including the effect of the QFT stress tensor as the star contracts. Notice here that we wish to retain the classical source as well, i.e. a fluid obeying the same equation of state as before. 

From this perspective a natural question to ask is the following. Consider again the FD trajectory \eqref{FD2} associated with classical gravitational collapse. What is the incoming stress tensor $P^{FD}_+(x^+)$ that has the FD trajectory \eqref{FD2} as its classical solution? For this will provide the `background' we want to perturb. The answer is of course obtained by plugging this solution into the classical eom, yielding:
\begin{align}
    P^{FD}_+(x^+)=\sqrt{\frac{\beta}{|x^+|} }.
\end{align}

This is the incoming stress tensor that has the FD orbit as its solution to the classical eom \eqref{eom:classical}. It diverges as we approach $x^+=0$, as it should: an infinite amount of energy is needed to accelerate a massive particle to the speed of light. This discussion suggests considering the more general class of incoming singular stress tensors with the asymptotic form
\begin{align}\label{singular:stress}
    P_+(x^+)= \frac{p}{(-x^+)^a},
\end{align}
for $x^+< 0$, where $p$ and $a$ are positive constants. The FD trajectory -- related to non-extremal black hole evaporation -- corresponds to $a=1/2$. 

Suppose that we fix this as the incoming data, but now solve the \textit{anomalous} equations of motion \eqref{aeom1} for the particle. Will the mirror again become null as it approaches the singular line $x^+\to 0$? This is the question we address next. We now proceed to solve the dynamics of \eqref{aeom2} with the incoming data set by \eqref{singular:stress}, again comparing the classical and quantum regimes.


\subsection{Classical}\label{subsec:Classical}
First, consider the solutions to the classical equations of motion \eqref{aeom3} with $q=0$ and incoming stress tensor given by \eqref{singular:stress}. The equation for the velocity $f'(x^+)$ is purely algebraic and we can immediately write down the solution,
\begin{align}\
    f'(x^+) = \left(A+\frac{p}{(-x^+)^a}\right)^2 , \ \ A \geq 0.
\label{classicalVel}
\end{align}
The constant $A$ must be non-negative, since for $x^+ \to -\infty$ we have $\sqrt{f'} \approx A$. Now, depending on the value of $a$, the solutions can be classified into three cases. As a common feature, the particle becomes null for any $a>0$. What differs among the solutions is the range of the coordinate $x^-$ and of proper time $\tau$ as $x^+$ approaches the singular line $x^+=0$. The latter are given by 
\begin{align}
x^-(x^+\to 0)=f(0) \approx \lim_{x^+ \to 0^-} \int_{x^+_i}^{x^+}dx^+ \left(\frac{p^2}{(-x^+)^{2a}}\right)
\end{align}
and
\begin{align}
    \tau(0)  \approx \lim_{x^+ \to 0^-} \int_{x^+_i}^{x^+}dx^+ \left(\frac{p}{(-x^+)^a}\right).
\end{align}
From these expressions, we can easily deduce the following. For $a<1/2$, the value of the other coordinate $x^-=f(x^+ \to 0)$ is finite, meaning the particle becomes null \textit{inside} the Penrose diagram (rather than asymptotically at its boundary $\mathcal I$). The proper time is finite obviously in this case. For $1/2 \leq a < 1$, the proper time is also finite, but since $x^-$ doesn't converge, the particle becomes null asymptotically as $x^- \to \infty$. For $a>1$ the particle also ends up at null infinity and proper time is infinite. These cases are illustrated in table \ref{tab}.
\begin{table}[!h]
\begin{center}
\begin{tabular}{|c|c|c|c|} 
 \hline
 $a \in$ & $(0,1/2)$  & $[1/2, 1)$ & $[1,\infty)$ \\
 \hline\hline
  $f(0)$ & finite & $\infty$ & $\infty$ \\
  \hline
  $f'(0)$ & $\infty$ & $\infty$ & $\infty$\\
  \hline
  $\tau(0)$ & finite & finite & $\infty$\\
  \hline
\end{tabular}
 \caption{Classification of the classical solutions, depending on the value of $a$: displayed are the coordinate $x^-=f(x^+)$, the velocity $f'(x^+)$ and proper time $\tau(x^+)$ as $x^+ \to 0$.}
 \label{tab}
 \end{center}
\end{table}\\
Since we are dealing with a purely classical field theory here, the reflected stress tensor is determined by its tensorial transformation law, in terms of $x^+$ 
\begin{align}
    T_{--}(x^+)=\frac{1}{(f'(x^+))^2}T_{++}(x^+)=\frac{m a p (-x^+)^{3 a-1}}{2 \left(A (-x^+)^a+p\right)^4},
\end{align}
where $T_{++}(x^+)=\frac{m}{2} P'(x^+)$. We see that in the limit $x^+ \to 0$, the stress tensor diverges for $a<1/3$, is equal to a constant for $a=1/3$, and vanishes for $a>1/3$.

\subsection{Quantum.}
After having discussed the response of the classical system to the singular stress tensor, we now turn to the quantum counterpart. To this end, we consider again \eqref{aeom3}, but now with $q > 0$. For convenience, let us restate the equation
\begin{equation}\label{quant:eq}
\sqrt{f'(x^+)} - \frac{q}{2}\frac{f''(x^+)}{f'(x^+)} - P_+ = A. 
\end{equation}
One of the main aims of this work is to compare the classical and quantum solutions with $A$, $p$, and $a$ held constant to the same values. In the gravitational context, this would be analogous to solving the field equations without/with quantum backreaction, for the same classical source. 

Consider first a general incoming momentum $P_+(x^+)$. We can write  equation \eqref{quant:eq} alternatively as
\begin{equation}\label{qeom_velocity}
    f''(x^+) = -\frac{2}{q} \left( (P_+(x^+) + A)f'(x^+) - f'(x^+)^{\frac{3}{2}} \right).
\end{equation}
Substituting $u = 1/\sqrt{f'}$ yields the following first order linear differential equation in $u$,
\begin{equation}
    u'(x^+) - \frac{P_+(x^+) + A}{q}u(x^+) + \frac{1}{q} = 0.
\end{equation}
We can readily write down a solution for the above equation with an integrating factor. It is given by
\begin{align}
\label{general:vel}
u(x^+) &= \frac{c_1-\mathcal{K}(x^+)}{\mathcal{I}(x^+)},
\end{align}
where $c_1$ is an integration constant and
\begin{align}
\mathcal{I}(x^+) &= \exp \left(-\frac{1}{q} \int (P(x^+)+A ) dx^+ \right)
\end{align}
and
\begin{align}
    \mathcal{K}(x^+)=\frac{1}{q}\int \mathcal{I}(x^+) dx^+.
\end{align}
Resubstituting yields the velocity
\begin{align}\label{full:quantum:sol}
   f'(x^+) = \frac{1}{u(x)^2}=\left(\frac{\mathcal{I}(x^+)}{c_1-\mathcal{K}(x^+)}\right)^2.
\end{align}

Equation \eqref{full:quantum:sol} is the implicit solution to the quantum back-reacted system for any incoming momentum. 

As motivated above, our interest lies in the case where the incoming momentum is given by \eqref{singular:stress}. Analytic expressions for $\mathcal{I}(x^+)$ are easily obtained, 
\begin{align} \label{intFactor}
\mathcal{I}(x^+)&=\exp \left(\frac{p (-x^+)^{1-a}}{q(1-a)}-\frac{A x^+}{q}\right), \hspace{0.5cm} \text{$a \neq 1$}\\
\mathcal{I}(x^+)&=(-x^+)^{p/q}\exp \left(-\frac{A x^+}{q}\right),\hspace{.95cm} \text{$a = 1$}.
\end{align}
The integral $\mathcal{K}(x^+)$ lacks an analytic form for general parameters $a$ and $A$ (but see the interesting case $a=1/2$ below).

Now regardless of the explicit form of $\mathcal K$, a comment on the choice of the integration constant $c_1$ and its physical interpretation is in order. Let $x^+_0<0$. If, for a given $c_1$ there exists an $x^+_0$ such that $c_1=\mathcal{K}(x^+_0)$, then the denominator of \eqref{full:quantum:sol} vanishes there and the particle becomes null. Now close to $x^+_0$ we can expand $\mathcal{K}(x^+)\approx \mathcal{K}(x^+_0)+(x^+-x^+_0) \mathcal{K}'(x^+_0)$, so it follows that
\begin{align}\label{sing:vacuum}
   f'(x^+) \approx \left(\frac{q}{x^+-x^+_0}\right)^2.
\end{align}
From \eqref{sing:vacuum} we see that these orbits correspond to the vacuum solutions \eqref{moeb} we observed before, where the particle approaches a constant proper acceleration.

The physical interpretation of \eqref{sing:vacuum} is as follows. In the previous section we found that, even if the incoming state is the vacuum, the quantum backreaction makes the particle runaway and become asymptotically null. For trajectories of type (i)-(iii), this occurs along a line of constant $x^+$. This tendency to become null doesn't go away once we introduce non-vanishing incoming energy. Indeed, if the incoming energy is very small or acts for a very short time (i.e. for certain choices of initial conditions or integration constants) the equation of motion will again be dominated by the vacuum dynamics and the particle will become null. This is what is happening in \eqref{sing:vacuum}. Since we have already examined the vacuum trajectories in detail above, in this section we wish to focus on the physics characterized by the incoming singular energy. Therefore we focus on those situations where the vacuum runaway \textit{does not} happen before we reach the line $x^+=0$ where the incoming stress tensor is divergent. This corresponds to choosing $x^+_0>0$.

For a power law behavior of the quantum velocity, the anomaly term in \eqref{quant:eq} to leading order scales as
\begin{align}
    \frac{f''}{f'} \sim \frac{1}{(-x^+)},
\end{align}
which means that for $a>1$ this term is sub-dominant compared to the incoming momentum $P_+$, implying that the leading order contribution of the quantum velocity is just the classical velocity
\begin{align}\label{null:orbits}
    f'(x^+) \approx  \frac{p^2}{(-x^+)^{2a}}.
\end{align}
Thus, in the following, we focus on $a<1$ and specifically $a=1/2$, where the anomaly term is dominant and qualitatively changes the solutions. 
\subsection{Fulling-Davies ($a=1/2$)}\label{subsec:a=1/2}
The main case of interest is the quantum backreacted solution for $a=1/2$, which is classically associated with the well-known FD trajectories \eqref{FD2} and the black hole horizon. In this case, $\mathcal{K}(x)$ has a simple closed-form expression given by
\begin{align}
    \mathcal{K}(x^+)=\frac{\mathcal{I}(x^+)}{A^{3/2}} \left(2 p \sqrt{q} F\left(\frac{p+A \sqrt{-x}}{\sqrt{A} \sqrt{q}}\right)-\sqrt{A} q\right),
\end{align}
where $F(x) = e^{-x^2} \int_0^x e^{y^2} dy$ is Dawson's Function. The velocity is then explicitly
\begin{equation}\label{full:a=1/2}
    f'(x^+) = \frac{A^2}{\left( 1- \frac{2 p}{\sqrt{Aq}}F\left(\frac{\sqrt{-x^+} A+p}{\sqrt{Aq}} \right) +A c_1 e^{\frac{A x^+ - 2 p \sqrt{-x^+}}{q}} \right){}^2}.
\end{equation}  
\begin{figure}[!h]
\includegraphics[scale=.9]{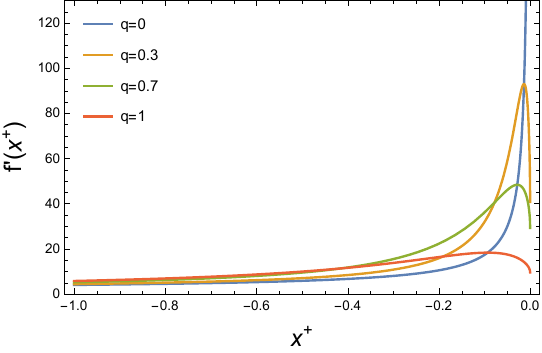}
\centering
\caption{Velocity $f'(x^+)$ for the classical ($q=0$) and the quantum equation of motion with incoming energy given by the Fulling-Davies one $(a=1/2)$. Here $A=1, c_1=0.4, p=1$. In the far past, the classical and quantum velocities match, whereas the behavior differs close to $x^+=0$, where the quantum solution remains finite.}
\label{full:a=1/2Graph}
\end{figure}

As explained above, \eqref{full:a=1/2} admits singularities, which are associated, however, with the vacuum orbits that have been previously examined.  Our interest lies now in exploring the new type of solutions, which are well-defined across the domain of $P_+$. These new solutions satisfy
\begin{align}
    c_1 > \mathcal{K}(0) = \frac{2 p }{A^{3/2} \sqrt{q}}F\left(\frac{p}{\sqrt{A} \sqrt{q}}\right)-\frac{1}{A}.
\end{align}
The reason for this is that the function $\mathcal{K}(x^+)$ is monotonically increasing, since $\mathcal{I}(x^+)>0$, and finite at zero. Thus if the above condition holds, $c_1>\mathcal{K}(x^+)$ and there are no poles in $f'(x^+)$.

In the far past, i.e. $x^+ \to -\infty$, the expansion of the velocity yields
\begin{align}
    f'(x^+) \approx \left(A+\frac{p}{\sqrt{-x^+}}\right)^2+\frac{p q}{(-x^+)^{3/2}}.
\end{align}
The first term corresponds to the classical velocity \eqref{classicalVel}, such that the classical and quantum velocities match in this limit. The second term is the first-order quantum contribution, which is manifestly positive: the quantum-backreacted particle accelerates faster initially than its classical counterpart.

In Fig. \ref{full:a=1/2Graph} we show for $a=1/2$ the classical velocity and also the quantum velocity for different values of $q$. While matching in the far past, 
a crucial difference between the classical and the quantum solution is the behavior close to $x^+=0$. Expanding \eqref{full:a=1/2} around $x^+=0$ leads to
\begin{equation}
    f'(x^+) \approx \xi \left( q + 4p \sqrt{-x^+}\right) + O\left((-x^+)^{3/2}\right),
    \label{taylor:a=1/2}
\end{equation}
with
\begin{equation}
    \xi = \frac{A^3}{\left(A^{3/2} c_1 \sqrt{q}-2 p F\left(\frac{p}{\sqrt{A} \sqrt{q}}\right)+\sqrt{A} \sqrt{q}\right){}^2}.
\end{equation}

\begin{figure}[!h]
\includegraphics[scale=1.8]{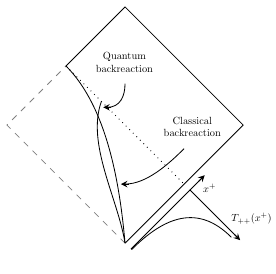}
\centering
\caption{Conformal diagram depicting the solutions to the equations of motion for the singular incoming momentum \eqref{singular:stress}, as shown in the right bottom of the diagram. Here $a=1/2$. The classical solution \eqref{classicalVel} and the quantum solution \eqref{full:a=1/2} closely resemble for a long time. However, while the classical solution becomes null and forms a horizon as the singular point is approached, the quantum solution deviates and the velocity remains finite.}
\label{singular}
\end{figure}

Thus, while the classical trajectory admits a `horizon' as $x^+ \to 0$ -- as reviewed in section \ref{subsec:Classical} --  the velocity for the quantum backreacted boundary particle remains finite. It is instead the acceleration that blows up in this limit. For these new orbits, that are not associated with the vacuum physics, the anomaly contribution prevents the particle from forming a `horizon', i.e. accelerating and becoming null at $\mathcal I$. This effect is depicted in Fig. \ref{singular}.

For this case, we will not consider the region beyond the singular line $x^+=0$. Our approach was to use the information about the classical gravitational collapse to determine the incoming stress tensor for the analogous mirror system, which led us to the FD-type sources. However, the endpoint of the classical trajectory determines the boundary of the black hole horizon, and as explained above the black hole interior is not contained in the mirror picture. Thus, the gravitational dynamics provides no information on how to continue beyond the singular line.

For the quantum velocity, there is exactly one turning point with $f''(x^+)=0$. This can be understood with a simple argument: At the turning point, the quantum equation of motion reduces to the algebraic classical one, such that the two graphs intersect. After the first intersection point, the quantum graph falls off, while the classical graph is monotonically increasing: there can't be another turning (intersection) point. Finding an analytic expression for the turning point for generic values of the integration constants is out of reach. 

For the specific case $A=0$ though, the turning point can be computed, since \eqref{full:a=1/2} simplifies to
\begin{align}\label{A=0:sol}
    f'(x^+) = \frac{4 p^4}{\left(-2 c_1 p^2 e^{-\frac{2 p \sqrt{-x^+}}{q}}-2 p \sqrt{-x^+}+q\right){}^2}\,.
\end{align}
For \eqref{A=0:sol} the turning point $f''(x^+_t)=0$ is given by
\begin{align}
  x^+_t=-\frac{q^2 \log ^2\left(\frac{2 c p^2}{q}\right)}{4 p^2}.
\end{align}
In the classical limit, $q \to 0$, this value goes to zero.  At the turning point the quantum crosses the classical graph and as the classical solution diverges as $x \to 0$, so must the value of the quantum velocity at the turning point.

\begin{figure}[!h]
\includegraphics[scale=.9]{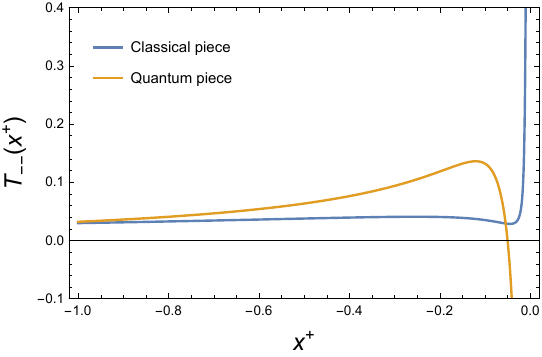}
\centering
\caption{Reflected stress tensor $T_{--}(x^+)$ for the quantum velocity \eqref{full:a=1/2}, consisting of the classically reflected incoming stress tensor (in blue) and the quantum piece associated with the conformal anomaly (in orange).  Here $A=1, c_1=1, p=1$, $q=1/10$.}
\label{FullReflectedPlot}
\end{figure}

As a final point, let us discuss the reflected stress tensor, which is determined by Eq. \eqref{reflection:quantum}. It consists of the Doppler-shifted incoming energy flux as well as the quantum anomaly piece. In Fig. \ref{FullReflectedPlot} we plot these two contributions for the velocity \eqref{full:a=1/2}. The leading order contribution of the anomaly piece for $x^+ \approx 0$ is given by
\begin{align}
    T_{--}(x^+) \sim -\frac{ 1}{(-x^+)^{3/2}},
\end{align}
which diverges negatively as $x^+ \to 0$, as can be seen in the Figure. 
\section{Averaged Null 
Energy.}\label{sec:ANEC}
Dynamical moving mirrors are also an interesting arena to study quantum energy inequalities in systems with boundaries. These inequalities are of central importance in general relativity, as they give restrictions on the stress energy tensor to prevent exotic phenomena like superluminal travel or traversable wormholes \cite{SuperluminalTravel, ThorneWormholes}. Sufficient for this is the null energy condition (NEC), $T_{\mu \nu}k^{\mu}k^{\nu} \geq 0$ with a null-vector $k$.  As is well known, in quantum theory such local energy conditions -- like the NEC -- are generically violated. The most prominent example is the aforementioned Casimir effect. Also, a moving mirror radiates according to \eqref{reflection:quantum}. In the prescribed setup, one can in principle create outgoing states with arbitrarily negative energy from the incoming vacuum.

A weaker condition class of constraints is given by `averaged' conditions along certain geodesics. Although less stringent, they can still prevent exotic phenomena from occurring. We consider the averaged null energy condition (ANEC),
\begin{align}\label{anec}
    \mathcal{E}=\int_{\gamma} d\lambda \ T^{tot}_{\mu \nu} k^{\mu} k^{\nu} (\gamma(\lambda)) \geq 0,
\end{align}
where the curve $\gamma$ is a null geodesic and $\lambda$ a affine parameter.

But if we are integrating over a null geodesic, what should we do when it reaches the boundary mirror? One proposal (in higher dimensions)\,\cite{Graham:2005cq} is that instead of the perfectly reflecting boundary, we instead solve the problem of a boundary containing a small hole through which the geodesic can travel and consider \eqref{anec} there. The calculation becomes more involved there, but the interpretation of the ANEC is straightforward.

Here we will take a different route. We will integrate the total stress tensor of the system -- including the fields and the boundary particle --  that we will derive below. As we will see, the contribution of the massive mirror will enter as a pure boundary term. 

Let us begin with the computation of the total stress tensor. We assume the existence of an action describing the system, 
\begin{align}
    S=S_{\text{CFT}} +S_p,
\end{align}
where the first part constitutes the matter CFT and the second part is the boundary particle. The stress tensor is then obtained in the usual way by taking the functional derivative of the action with respect to the metric $g$,
\begin{align}
    T^{tot}_{\mu \nu}=\frac{2}{\sqrt{-g}}\frac{\delta S}{\delta g^{\mu \nu}}.
\end{align}
The action of the particle is given by
\begin{align}
    S_p&=-m \int d^2x \int d\sigma \sqrt{g_{\mu \nu}\dot{x}^{\mu}\dot{x}^{\nu}}\delta^2(x-x(\sigma)),
\end{align}
where $\sigma$ is an arbitrary parameter describing the trajectory and the dot refers to the derivative with respect to this parameter. The explicit form of $S_{\text{CFT}}$ is not relevant to the discussion here. Varying the total action with respect to the metric gives the total stress tensor
\begin{align}\label{totalstress}
    T^{tot}_{\mu \nu}=&T_{\mu \nu}+ \frac{m }{\sqrt{-g}}  \int d\sigma \left(\frac{\dot{x}_\mu \dot{x}_\nu}{\sqrt{\dot{x}^\sigma \dot{x}_\sigma}}  \delta^2(x-x(\sigma))\right),
\end{align}
where by definition
\begin{align}
    T_{\mu \nu}=\frac{2}{\sqrt{-g}}\frac{\delta S_{\text{CFT}}}{\delta g^{\mu \nu}}.
\end{align}
Evaluating for the flat metric and replacing the parameter $\sigma$ with proper time gives
\begin{align}
    T^{tot}_{\pm \pm}=T_{\pm \pm}+ \frac{m}{2} v^{\mp} \delta(x^{\pm}-x^{\pm}(\tau)),
\end{align}
with $v^{\mp} = dx^{\mp}/d\tau$ as before. It is this quantity for which we evaluate the ANEC in \eqref{anec}.
In our setup, we have two different null lines along which we can compute the averaged energy. First, keeping $x^-$ constant and integrating along $x^+$,
\begin{align}\label{ANEC++}
    \mathcal{E}_+(\tau)=\int_{x^+(\tau)}^{\infty} T_{++} d\tilde{x}^+ + \frac{m}{2}v^-(\tau),
\end{align}
and second, keeping $x^+$ constant and integrating along $x^-$,
\begin{align}\label{ANEC--}
    \mathcal{E}_-(\tau)=\int_{-\infty}^{x^-(\tau)} T_{--} d\tilde{x}^- + \frac{m}{2}v^+(\tau).
\end{align}
Note that the boundary contribution to the averaged energy is always positive for future directed curves, whereas the field term can be positive or negative depending on the energy flux $T_{\pm \pm}$. 

\begin{figure}[!h]
\includegraphics[scale=1.5]{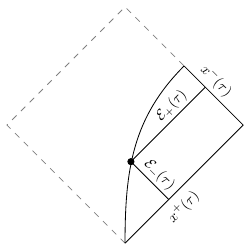}
\centering
\caption{Illustration of the integration paths for the averaged null energy $\mathcal{E}_{\pm}$, here for the trajectory (iv) of the vacuum solutions. For a given $\tau$, $\mathcal{E}_{+}(\tau)$ is computed along a line of fixed $x^-(\tau)$; $\mathcal{E}_{-}(\tau)$ is computed along a line of fixed $x^+(\tau)$. Both integration paths pick up a contribution of the boundary particle.}
\label{path-for-ANEC}
\end{figure}

Some comments are in order. First, the limits in the integral of $T_{\pm\pm}$ reflect the fact that all matter fields are restricted to exist only on the right side of the mirror, see Fig. \ref{path-for-ANEC} for the integration paths. Second, the averaging above is performed along null lines that intersect the worldline of the boundary particle, such that the coordinates $x^{\pm}$ to specify the integration paths are functions of proper time via the map $x^{\pm}(\tau)$ and thus also $\mathcal{E}_{\pm}$.

Using the equations of motion \eqref{aeom2}, we can eliminate the integral of the stress tensor to obtain a simple expression for the ANEC in terms of the $\dot v^\pm$ as
\begin{align}
    \mathcal{E}_+(\tau)&=\frac{m}{2} \left( A^- + q \dot{v}^- \right)\label{ANEC--Va},\\
    \mathcal{E}_-(\tau)&=\frac{m}{2} \left( A^+ - q \dot{v}^+ \right).\label{ANEC--VaMinus}
\end{align}
Remarkably, the ANEC --  which is in general hard to deal with given its non-locality -- has become purely a boundary term. Next, we will study its properties for some of the situations studied above.

\subsection{Examples.}
Let us begin with the vacuum solutions studied in section \ref{sec:vacuum}. As $T_{++}=0$, the averaged energy along $x^+$ reduces to the manifestly positive contribution of the particle and the ANEC is trivially satisfied. 

More interesting is the averaged energy along $x^-$. Regarding the energy flux of the vacuum solutions, cases (i) and (ii) have $T_{--} > 0$, i.e. the stronger NEC is satisfied and consequently the ANEC. On the other hand, cases (iii) and (iv) have $T_{--}<0$ everywhere. For these cases, it is not immediately clear from \eqref{ANEC--} how the averaged energy behaves.

When evaluated for the vacuum solutions \eqref{vac:vel}, the averaged energy \eqref{ANEC--VaMinus} takes the form
\begin{align}
    \mathcal{E}_-(\tau)=\frac{B m e^{\tau/q}}{2\left(A+B e^{\tau/q}\right)^2}+A^+.
\end{align}
The integration constant $A^+$ relates to $A$  as follows
\begin{align}
    A^+=\frac{A}{(v^-(\tau_i))^2},
    \label{A+ constant}
\end{align}
where $x^-(\tau_i)$ is the lower boundary of the integration in \eqref{ANEC--}. 
For trajectory (iii), where $A<0,B>0$, the first term is positive, but $A^+$ as given by the expression \eqref{A+ constant} is negative and singular, since the velocity $v^-(\tau)$ becomes zero as $\tau \to \tau_c^+$. For trajectory (iv), where $A>0, B<0$, $A^+$ is finite but the first term blows up negatively as proper time approaches the critical time. Thus, the averaged energy is negatively divergent in both these cases.

\begin{figure}[!h]
\includegraphics[scale=.9]{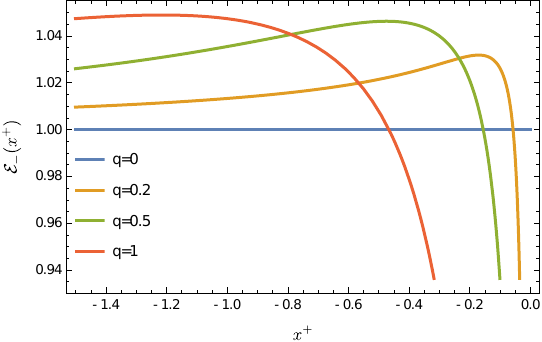}
\centering
\caption{Averaged null energy, $\mathcal{E}_-(x^+)$ computed for the solution given by Eq. \eqref{full:a=1/2}. Here $A=1, c_1=1, p=1$. Case $q=0$ corresponds to the classical backreaction.}
\label{ANEC-for-FD}
\end{figure}

Now more generally, \eqref{ANEC--Va} tells us that $\dot v^+(\tau_c) \to \infty$ is a necessary and sufficient condition for $\mathcal{E}_-$ to diverge negatively.
For the vacuum orbits, the trajectories that lead to $\mathcal{E}_- \to -\infty$ have the property that they become null at some finite proper time, $\tau_c$, which are only a subset of cases for which $\dot{v}^+ \to \infty$. Interestingly, the non-vacuum solution discussed in \ref{subsec:a=1/2}, given by \eqref{full:a=1/2} corresponds to the case where the velocity remains finite but $\dot v^+ \to \infty$. To see this, we first convert the expression given by \eqref{ANEC--VaMinus} to the different parameterization, namely $x^- = f(x^+)$, which reads:
\begin{align}\label{ANEC--qback}
     \mathcal{E}_-(x^+) = \frac{m}{2} \left( A^+ + \frac{q}{2}\frac{f''(x^+)}{(f'(x^+))^2} \right).
\end{align}
Now, as $x^+ \to 0^-$, using the series expansion of $f'(x^+)$, given by \eqref{taylor:a=1/2}, we get $f''(x^+)/(f'(x^+))^2$ goes to minus infinity and so does the ANEC, as shown in Fig. \ref{ANEC-for-FD}. 

We leave as future work a more thorough exploration of the ANEC for systems with boundaries. 




\section{Summary and outlook.}\label{sec:summary}

The interaction of moving mirrors and QFTs has been extensively studied in connection to black holes and the Dynamical Casimir effect. The usual approach is to consider predetermined trajectories that are chosen to reproduce either of these effects by construction. 

In this paper, we have investigated the dynamics of a massive relativistic moving mirror acting as a perfectly reflecting boundary of a CFT$_2$. Instead of following a predetermined trajectory, the boundary particle obeys its own equation of motion, $F^\mu=ma^\mu$, and thus the problem is fully dynamical.

Our goal was to study the effect of the quantum stress tensor -- via the 2d conformal anomaly -- on the particle's trajectory, analogous to the challenges faced in General Relativity when incorporating the stress tensor of quantum fields into the semi-classical Einstein's equation during gravitational collapse. While being difficult in the latter context, the simplicity of the moving mirror model allows for a much more analytic approach. 

Our main results are the following. Upon quantization of the field, the backreaction produced by the conformal anomaly emerges as a third-order term in the equations of motion.  If the incoming quantum state is the vacuum, the massive particle suffers from a runaway (similar to those familiar in electrodynamics), eventually becoming asymptotically null, see Fig. \ref{VacuumPenrose}. Interestingly, the asymptotic trajectories have constant proper acceleration and although they produce no reflected radiation, they do have an associated horizon. In fact as we saw in section \ref{sec:vacuum}, these vacuum solutions map to extremal black holes of mass $M=\frac{c}{24\pi} \frac{1}{G m}$.

Next, in order to mimic the spacetimes of gravitational collapse, we focused on singular incoming stress tensors of the form $P_+(x^+) \sim  (-x^+)^{-a}$ that produce orbits with a horizon when the field remains classical. We found that for $a<1$, the additional term accounting for the quantum backreaction qualitatively alters the solutions, preventing the formation of an acceleration horizon. In particular, this includes the case $a=1/2$, where the solutions of the classical system are the Fulling-Davies orbits corresponding to gravitational collapse. 

Finally we considered the Averaged Null Energy Condition (ANEC). Here we introduced a novel concept: for a $1+1$-dimensional spacetime with a dynamical boundary, we defined the ANEC integral by including the stress tensor of the field \textit{plus} that of the boundary particle. Via the particle's equations of motion, the integral over the field stress tensor is immediate, rendering the entire ANEC operator a pure boundary term. We explored its properties for some of the solutions presented above.

Future work may include several directions. Here we considered the case where the fields are restricted to the right side of the mirror, while the left component of the spacetime doesn't exist. Thus we could examine the dynamics with both sides included. On the other hand, we restricted to considering a classical (non-quantum) boundary particle. Quantizing the dynamics of the particle would lead to a fully quantum system. It would also be interesting to understand the role of the ANEC boundary terms for the standard Casimir effect.

\section{Acknowledgements}
The authors would like to thank Michael R.R. Good for insightful comments on the draft.

\bibliography{Mirror_refs}
\appendix
\end{document}